\documentclass[useAMS, usenatbib, usegraphicx]{mn2e}
\usepackage{times}

\title[The outer Galaxy at 1420 and 408 MHz]{A sharper view of the
  outer Galaxy at 1420 and 408 MHz from the Canadian Galactic Plane
  Survey I: Revisiting the KR catalogue and new Gigahertz Peaked
  Spectrum sources}

\author[C. R. Kerton]{C. R. Kerton,$^1$\thanks{E-mail: kerton@iastate.edu}\\
$^1$Iowa State University, Department of Physics and Astronomy, Ames,
IA, 50011, USA}

\begin{document}

\maketitle

\begin{abstract}
Arcminute-resolution radio continuum images at 408 and
1420 MHz from the Canadian Galactic Plane Survey (CGPS) have been used
to reexamine radio sources listed in the \citet{kal80}
catalogue. This catalogue is of particular interest to Galactic
studies as it lists both extended and compact radio sources found in
the second Galactic quadrant. We have determined the nature (extended
vs. compact, Galactic vs. extragalactic) of all of these bright radio
sources. A number of large \mbox{H\,{\sc ii}} regions with no optical
counterparts are highlighted along with a sample of large radio
galaxies. Many sources previously thought to be extended Galactic
objects are shown to be point sources. A sample of point sources with
flat or rising spectra between 408 and 1420 MHz has been compiled, and
within this sample likely Gigahertz Peaked Spectrum sources have been 
identified.
\end{abstract}


\section{Introduction} \label{sec:intro}

Radio continuum observations of the second quadrant of our Galaxy
($90\degr < l < 180\degr$) provide an unmatched opportunity for
studying the structure and content of a spiral arm in detail. The
Perseus Arm dominates Galactic structure in this quadrant and is
viewed almost perpendicular to its long axis over the entire longitude
range. The more distant Outer Arm is also well placed for study in
this quadrant and in both cases confusion from Local Arm sources is
minimal (cf. the view of the Galaxy around $l\sim
75\degr$ looking along the Local Arm). The previously best view of
this region in the radio continuum (at 1420 MHz) was a series of
surveys done by the Effelsberg 100-m telescope at 9-arcmin
resolution. The surveys were summarized in the \citet{kal80} and
\citet*{rrf97} catalogues (KR and RRF respectively). RRF
provides a listing of small diameter sources ($<16$ arcmin in extent)
with an 80 mJy flux density limit (for point sources).  The KR
catalogue has a higher flux density limit (0.3 Jy) but is of
particular interest to Galactic studies as it lists both compact and
extended objects.

The new Canadian Galactic Plane Survey (CGPS; \citealt{tay03}) data provide an
unprecedented view of the continuum radiation at both 1420 and 408 MHz
from the outer Galaxy. The data have arcminute-scale resolution and have
full spatial frequency sensitivity crucial for the detection of extended
structures.

In this paper we first revisit the sources found in the KR catalogue.  
\citet{fic86} obtained high resolution VLA images of the sources originally
classified as point sources in KR. For these sources we are primarily
interested in observing the few of them that had poor VLA
observations and to look for inverted spectrum sources.  \citet{tru90}
obtained one-dimensional scans at 7.6 and 31.3 cm of most of the
extended KR sources using the RATAN-600 telescope and found that many
of the apparently extended KR objects were compact sources ($\leq$
1-arcmin scale). \citet{tru90} also suggested that a number of the KR
objects were previously unknown compact Galactic supernova remnants
(SNRs).  We have reexamined all of these sources using the higher
resolution and regular beamshape of the CGPS data and have been able to better
determine the nature of all of the extended KR objects.

In the course of this study a new
sample of extragalactic Gigahertz Peaked Spectrum (GPS) sources has
been compiled. CGPS data have also revealed numerous new extended
emission features in the second quadrant including both low-surface
brightness extended emission and narrow filamentary features -- both
of which tend to be missed in the lower resolution surveys. The second
paper in this series will present a complete catalogue of all extended
emission features seen in the CGPS radio continuum data thus providing
an updated version of the comprehensive catalogue compiled by \citet{fic86}. 

In the next section we review the properties of the CGPS 1420 and 408
MHz data. In Sections~\ref{sec:kr} and \ref{sec:kr-p} the CGPS view of
the KR sources is presented. Flat and inverted spectrum sources are
discussed in Section~\ref{sec:fiss} and conclusions are presented in
Section~\ref{sec:conc}.

\section{Observations} \label{sec:observe}

The goal of the CGPS is to enhance the study of our Galaxy by
obtaining arcminute-resolution images of all of the major components of
the interstellar medium (ISM) in our Galaxy. Radio continuum
observations made as part of this project were obtained using the
seven-element interferometer at the Dominion Radio Astrophysical
Observatory (DRAO) in Penticton, Canada \citep{lan00}. Details of the
CGPS radio continuum observations, data reduction and data
distribution are discussed at length in \citet{tay03}. CGPS
observations currently cover $65\degr < l < 175\degr$ between $-3\fdg5
< b < +5\fdg5$ encompassing almost the entire second
quadrant. The 1420 MHz observations have a nominal 1-arcmin
resolution and both the 1420 and 408 MHz survey images were
constructed with full spatial frequency coverage by combining the
interferometer data with data from surveys using the Effelsberg
single-dish and the Stockert single-dish telescopes. This provides
sensitivity to extended structure which is very important for Galactic studies.

The simultaneous 408 MHz images, with nominal 3-arcmin resolution,
provide invaluable data on the shape of the radio continuum spectrum
as parameterized by the spectral index ($\alpha_{408}^{1420}$) between
408 and 1420 MHz (where flux density F$_\nu \propto \nu^\alpha$). In
this paper we refer to inverted-spectrum sources as those with
$\alpha_{408}^{1420} \geq +0.25$ and flat-spectrum sources as those
with $|\alpha_{408}^{1420}| < 0.25$.

We also make use of the Mid-infrared Galaxy Atlas (MIGA;
\citealt{ker00}) and Infrared Galaxy Atlas (IGA; \citealt{cao97})
arcminute resolution infrared images which make up part of the larger
CGPS data collection. These infrared images are very useful in the
identification of Galactic \mbox{H\,{\sc ii}} regions in cases where
there is no associated optical emission or available radio recombination line
observations.

Flux density measurements were made using software contained in the DRAO Export
Software Package.  Point source flux densities were obtained using the
``fluxfit'' program which fits Gaussians to the image and makes use of
the beam shape information available in the CGPS data. Extended
sources were measured using the ``imview'' program which allows the
user to interactively derive background levels to use in determining
the flux densities.

\section{Extended sources in the KR catalogue} \label{sec:kr}

The KR catalogue is based on 1420 MHz radio continuum observations made
at 9-arcmin resolution with the Effelsburg 100-m
telescope. \citet{kal80} identified 236 radio sources with flux density
$F_\nu > 0.3$ Jy including point sources and extended objects up to
30-arcmin in diameter.  The catalogue covered $l=93\degr$ to
$l=162\degr$ and $|b| < 4\degr$.  Extended sources were subdivided
into three categories depending upon their apparent size: EP
(partially extended), E (extended) and VE (very extended). EP sources
had a greatest extent of $<$~9-arcmin, E sources had greatest extents
between 11-arcmin and 20-arcmin, while VE sources had greatest extents
between 20-arcmin and 30-arcmin.  

\begin{figure}
\includegraphics[width=84mm]{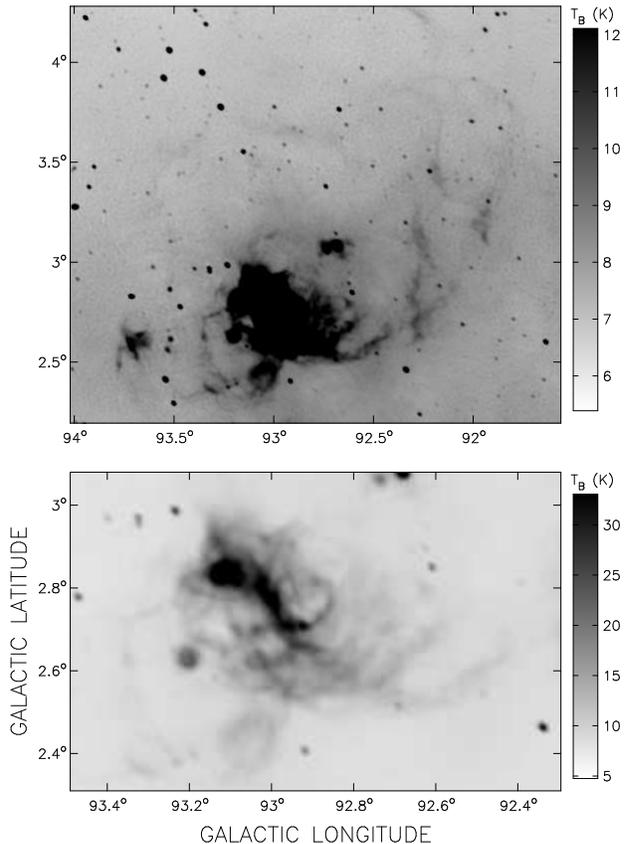}
%\vspace{3.5cm}
\caption{1420 MHz images of KR~1, an enormous \mbox{H\,{\sc ii}}
  region in the Perseus Arm.  The top panel shows the full extent of
  the region including extensive filamentary structure seen between
  $l=92\degr$ and $l=92\fdg5$. KR~4 is located in the lower left
  corner of this panel around $l=93\fdg75$. The lower panel shows the
  central region and reveals an intricate combination of filaments and 
  bubble-like structures.}
\label{fig:kr1}
\end{figure}

\begin{figure*}
\centering
\begin{minipage}{140mm}
\includegraphics[width=140mm]{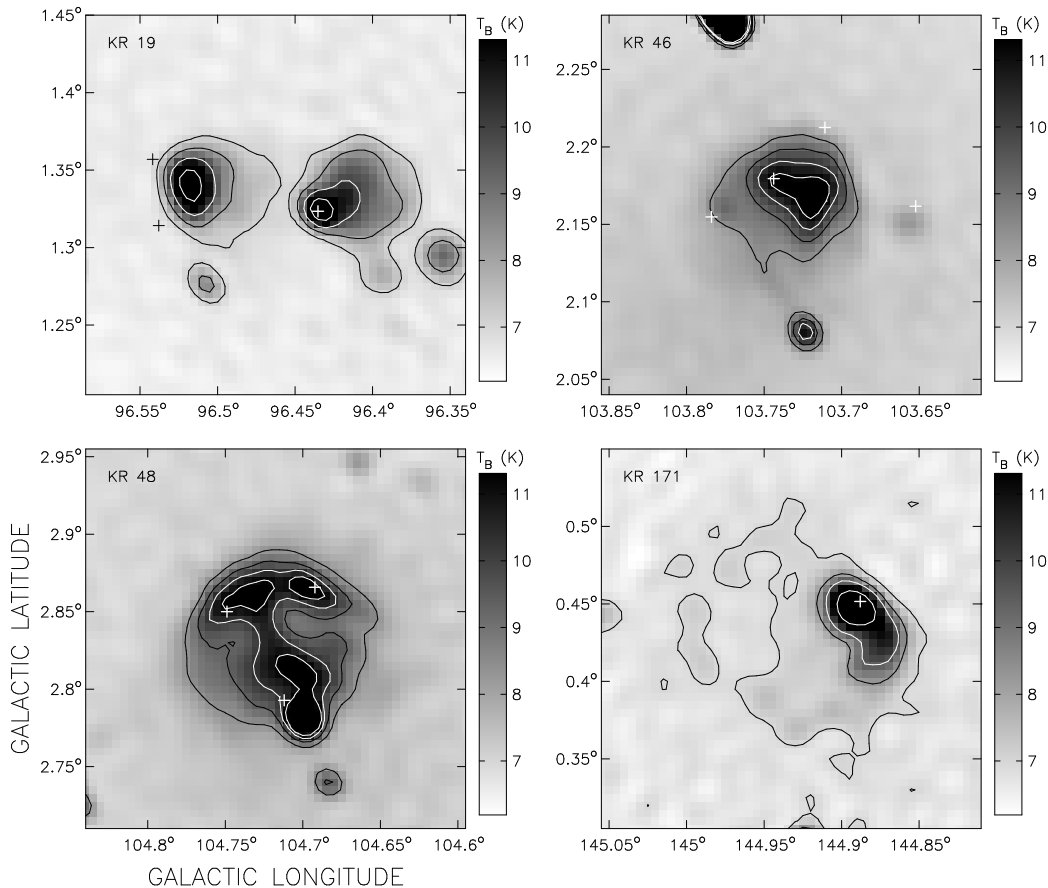}
%\vspace{3.5cm}
\caption{Small extended regions at 1420 MHz. The KR designation is
  given in each panel and the crosses indicate the positions of bright
  \emph{IRAS} point sources. All of these objects are Galactic
  \mbox{H\,{\sc ii}} regions. Contours for KR 19 are at 7, 8, 10 and
  13~K; for KR 46 and KR 48 at 8, 9, 10 and 11~K; and for KR 171 at
  5.5, 6.5, 7.5 and 10~K.}
\label{fig:small_ex}
\end{minipage}
\end{figure*}
 
\subsection{Very-extended (VE) sources} \label{sec:kr-ve}

Data on the twelve very-extended (VE) objects identified by
\citet{kal80} are listed in Table~\ref{tab:ve}.  The first column
gives the KR catalogue number. Letters following the KR number are
used in cases where the object is actually a multiple source at
arcminute resolution and are not part of the original classification
(e.g., KR206A). Columns 2 through 5 give the flux density measurements
and 1$\sigma$ error estimates at 1420 and 408 MHz from the CGPS
data. The spectral index between 408 and 1420 MHz
($\alpha_{408}^{1420}$) is given in column 6 followed by the angular
scale of the source as seen in the 1420 MHz images in column 7.
The final column provides extra information about the source, such as
an association with well-known optically visible \mbox{H\,{\sc ii}}
region or SNR. For extended (at 1-arcmin resolution) sources the RRF
catalogue number is given if applicable, and for all of the
arcminute-scale point sources the NRAO VLA Sky Survey (NVSS;
\citealt{con98}) catalogue designation is provided.    

%Table of Sources originally classified as VE in Kallas & Reich 1980

\begin{table*}
\centering
\begin{minipage}{140mm}
\caption{1420 MHz and 408 MHz data for VE sources}
\label{tab:ve}
\begin{tabular}{lccccccl}
\hline
KR & F$_\nu$ (1420) & $\sigma$ (1420) & F$_\nu$ (408)    & $\sigma$ (408) & $\alpha_{408}^{1420}$ & Diameter & Notes \\
   &   (mJy)        &       (mJy)     &    (mJy)         &     (mJy)      &                       &  \arcmin &       \\
\hline
1    & $3.26\times10^4$ &  $9.9\times10^2$  &  $3.35\times10^4$  &  $7.9\times10^2$ &  $-0.02$   & 120  & RRF 861; \mbox{H\,{\sc ii}} Region \\
3    & $4.48\times10^3$ &  $1.0\times10^2$  &  $4.29\times10^3$  &  $2.5\times10^2$ &  $+0.03$   &  18  & RRF 863; \mbox{H\,{\sc ii}} Region \\
6    & $7.92\times10^2$ &  $5.0\times10^1$  &  $5.01\times10^2$  &  $2.9\times10^1$ &  $+0.4$    &  12  & \mbox{H\,{\sc ii}} Region          \\
20   & $1.01\times10^3$ &  $5.7\times10^1$  &  $9.68\times10^2$  &  $1.1\times10^2$ &  $+0.03$   &  15  & \mbox{H\,{\sc ii}} Region          \\
47   & $2.99\times10^3$ &  $1.3\times10^1$  &  $2.08\times10^3$  &  $7.5\times10^1$ &  $+0.3$    &  20  & Sh 2-135            \\
65   & $1.10\times10^3$ &  $5.4\times10^1$  &  $9.68\times10^2$  &  $2.6\times10^2$ &  $+0.1$    &  12  & Sh 2-151            \\
122  & $6.43\times10^2$ &  $3.9\times10^1$  &  $4.36\times10^2$  &  $1.8\times10^1$ &  $+0.3$    &  24  & \mbox{H\,{\sc ii}} Region          \\
166A & $7.35\times10^3$ &  $2.2\times10^2$  &  $1.52\times10^4$  &  $4.6\times10^2$ &  $-0.6$    &   1  & NVSS J032719+552029 \\
166B & $1.23\times10^3$ &  $3.9\times10^1$  &  $2.77\times10^3$  &  $8.4\times10^1$ &  $-0.7$    &   1  & NVSS J032744+552226 \\
175A & $2.31\times10^3$ &  $7.0\times10^1$  &  $4.86\times10^3$  &  $1.5\times10^2$ &  $-0.6$    &   1  & NVSS J032952+533236 \\
175B & $7.45\times10^1$ &  $5.3\times10^0$  &  $1.51\times10^2$  &  $4.5\times10^0$ &  $-0.6$    &   1  & NVSS J033003+532944 \\
180  & $4.5\times10^ 2$ &  $1.4\times10^1$  &  $1.03\times10^3$  &  $3.1\times10^1$ &  $-0.7$    &   1  & NVSS J035927+571706 \\
206A & $3.37\times10^2$ &  $1.0\times10^1$  &  $4.96\times10^2$  &  $1.5\times10^1$ &  $-0.3$    &   1  & NVSS J043523+511422 \\
206B & $2.28\times10^2$ &  $6.8\times10^0$  &  $1.08\times10^2$  &  $3.2\times10^0$ &  $+0.6$    &   1  & NVSS J043621+511253 \\
210A & $1.84\times10^2$ &  $5.6\times10^0$  &  $5.22\times10^2$  &  $1.6\times10^1$ &  $-0.8$    &   1  & NVSS J043342+502428 \\
210B & $7.89\times10^1$ &  $2.7\times10^0$  &  $1.52\times10^2$  &  $6.3\times10^0$ &  $-0.5$    &   1  & NVSS J043357+502420 \\ 
\hline
\end{tabular}
\end{minipage}
\end{table*}

Seven of these sources are Galactic \mbox{H\,{\sc ii}} regions. These
sources all have flat or inverted spectral indices and have extensive infrared
emission visible in the \emph{IRAS} images. Five of the  \mbox{H\,{\sc ii}}
regions have no optical counterparts.  KR~1 is an enormous
\mbox{H\,{\sc ii}} region stretching up to 2\degr in size (see
Figure~\ref{fig:kr1}). Radio recombination line emission has been
detected from the region at V$_\mathrm{LSR} \sim -60$~km~s$^{-1}$
\citep{fic86} yielding a kinematic distance (accounting for known
streaming motions) of $\sim 4.5$ kpc, which implies that the region is
also physically large ($\sim 200$ pc). Note that the RRF 861 source
associated with the region refers only to a compact source making up only
a small portion of this extensive region.

KR~3, often incorrectly classified as a SNR, is a Galactic
\mbox{H\,{\sc ii}} region with a blister morphology which was
extensively studied by \citet{fos01}. In addition to
the flat radio spectrum and extensive associated infrared emission,
radio recombination line emission from the region has also been
detected \citep{fos01} solidifying its classification as an
\mbox{H\,{\sc ii}} region. RRF 863 is centered on the bright radio emission
associated with the \mbox{H\,{\sc ii}} region/molecular cloud
interface while the entire region extends up to 0\fdg3 in size.

KR 6, KR 20 and KR 122 are all classified as extended Galactic
\mbox{H\,{\sc ii}} regions on the basis of their radio spectrum and
associated infrared emission. None of these regions have known optical
counterparts. Finally there are two radio sources associated with
optically visible \mbox{H\,{\sc ii}} regions.  KR 47 is radio
emission, about 20-arcmin in extent, associated with the Sh 2-135
\mbox{H\,{\sc ii}} region, while KR 65 is diffuse radio emission, about
12-arcmin in extent, that is apparently associated with Sh 2-151. 

The remaining five VE sources turn out to be point sources at
arcmin-scale resolution.  KR 180 appears to have been misclassified
because of nearby diffuse radio emission associated with Sh 2-214.
This object was also listed by \citet{tru90} as being extended and
being a possible SNR but the CGPS data show this is not the case.
The other sources tend to be pairs of point sources with
separations $<$9-arcmin. All but one of the point sources have
a non-thermal spectral index and no detectable infrared emission,
consistent with them being distant extragalactic
objects.  The exception is the compact massive star-forming region KR
206B (NVSS J043621+511254) which has an inverted spectrum ($\alpha =
+0.6$) and is associated with the bright infrared source
IRAS~04324+5106 (RAFGL 5124).

\subsection{Extended (E) sources} \label{sec:kr-e}

\citet{kal80} listed 48 of these sources. Table~\ref{tab:e}
summarizes the CGPS view of this sample using the same notation as in
Table~\ref{tab:ve}.  Note that KR 86 was not observed in the CGPS
and KR 35 is apparently a spurious source; no bright point source or
region of diffuse emission was detected near its catalogued position.

%Table of Sources originally classified as E in Kallas & Reich 1980

\begin{table*}
\centering
\begin{minipage}{140mm}
\caption{1420 MHz and 408 MHz data for E sources}
\label{tab:e}
\begin{tabular}{lccccccl}
\hline
KR & F$_\nu$ (1420) & $\sigma$ (1420) & F$_\nu$ (408)    & $\sigma$ (408) & $\alpha_{408}^{1420}$ & Diameter & Notes \\
   &   (mJy)        &       (mJy)     &    (mJy)         &     (mJy)      &                       &  \arcmin &       \\
\hline
4    & $1.06\times10^3$ &  $3.7\times10^1$  &  $9.01\times10^2$  &  $1.7\times10^1$ &  $+0.1$   & 12  & RRF 865; \mbox{H\,{\sc ii}} Region \\
7    & $2.69\times10^3$ &  $8.1\times10^1$  &  $2.45\times10^3$  &  $7.3\times10^2$ &  $+0.07$  & 12  & RRF 874; \mbox{H\,{\sc ii}} Region \\
19A  & $1.70\times10^2$ &  $7.3\times10^0$  &  $7.85\times10^2$  &  $3.6\times10^0$ &  $+0.1$   &  5  & RRF 903; \mbox{H\,{\sc ii}} Region \\
19B  & $1.47\times10^2$ &  $3.4\times10^0$  &  $1.29\times10^2$  &  $3.9\times10^0$ &  $+0.1$   &  4  & RRF 903; \mbox{H\,{\sc ii}} Region \\
21A  & $3.78\times10^2$ &  $1.1\times10^1$  &  $9.09\times10^2$  &  $2.7\times10^1$ &  $-0.7$   &  1  & NVSS J214343+523958 \\
21B  & $3.64\times10^2$ &  $1.1\times10^1$  &  $1.05\times10^2$  &  $3.1\times10^1$ &  $-0.8$   &  1  & NVSS J214418+524501 \\
\hline
\end{tabular}
\medskip
Table~\ref{tab:e} is presented in its entirety in the electronic
edition of the journal.
\end{minipage}
\end{table*}

One source, KR 196, is a very large ($\sim$25-arcmin diameter) region of
bright radio emission associated with the optical \mbox{H\,{\sc ii}} region Sh
2-206. Seven other sources match the original classification
(diameters between 11-arcmin and 20-arcmin). Three of these (KR 55, 91
and 98) are associated with radio emission from known optical
\mbox{H\,{\sc ii}} regions, while three others (KR 4, 7, and 80) are
\mbox{H\,{\sc ii}} regions with no optical counterparts.  All of these
objects have flat or inverted radio spectra and have associated
infrared emission. Finally KR 101 is the well-studied SNR 3C 10 (Tycho's SNR).

Five other regions (KR 19, 46, 48, 171 and 198) are smaller extended
regions. KR 19 consists of two compact \mbox{H\,{\sc ii}} regions with
the western (19A) region being associated with IRAS
21336+5333 and the eastern one (19B) being associated with two
infrared sources IRAS 21340+5339 and IRAS 21340+5337 (see
Figure~\ref{fig:small_ex}).  KR 46 is a compact \mbox{H\,{\sc ii}}
region that shows hints of a blister morphology at 1-arcmin
resolution. The radio spectrum is thermal and there is bright infrared
emission associated with the region.  \citet{tru90} suggested that KR
48 and KR 171 were possible Galactic supernova remnants. However the
CGPS data show the regions have inverted (KR 48) and flat (KR 171)
radio spectra and are associated with bright diffuse infrared emission
and IRAS point sources. Thus
it is more likely that they are both Galactic \mbox{H\,{\sc ii}}
regions.  Finally KR 198 is associated with the optical
\mbox{H\,{\sc ii}} region Sh 2-207.

KR 168 consists of two slightly elongated sources separated by $\sim
4.5$ arcmin. It is likely that these sources are extragalactic
jets that are just barely resolved at 1-arcmin resolution. It is
not clear that the two sources are physically associated.  KR 188 also
consists of two elongated sources with a similar point source plus faint jet
structure with the point sources being separated by $\sim$4
arcmin. In this case the two objects do share common diffuse emission
and the jet structures both point back to a common point suggesting
that they are physically related.  In Table~\ref{tab:e} the NVSS designations
for the point-like portions of these objects are given.

\begin{figure}
\includegraphics[width=84mm]{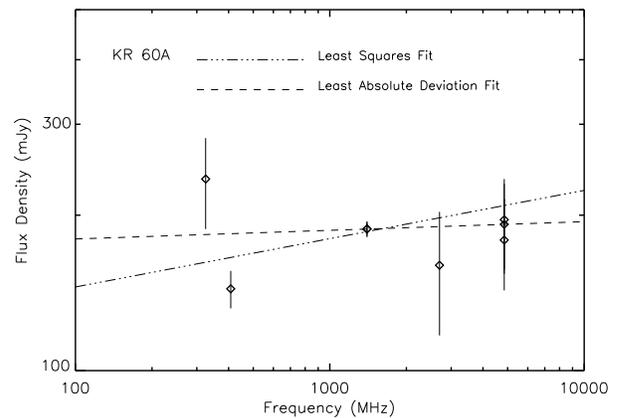}
%\vspace{3.5cm}
\caption{KR 60A, a flat-spectrum radio galaxy. CGPS data are at
  408 and 1420 MHz. Other data points were obtained from \citet{vol05}.}
\label{fig:kr60a}
\end{figure}

The remaining ``extended'' KR sources are all actually point
sources at 1-arcmin resolution. The majority of these sources are
extragalactic as they have strongly non-thermal spectral indices, are
unresolved at 1-arcmin resolution, and have no associated infrared
emission.  Three of the sources have flat spectra (KR 63, 189 and
192A) and two have inverted spectra (KR 53 and 60A).  None of the flat
spectrum sources have associated infrared emission and, given that they
all have $\alpha_{408}^{1420} = -0.2$, they are also most likely
extragalactic objects.  KR 53 is associated with the optical \mbox{H\,{\sc ii}}
region Sh 2-138.  Finally, KR 60A is apparently a flat-spectrum radio
galaxy. There is no associated infrared emission and, combining the
CGPS flux density measurements with data obtained using
SPECFIND \citep{vol05}, we find a very flat spectral
index of $+0.09\pm0.05$ over the range from 325 to 4800 MHz as
illustrated in Figure~\ref{fig:kr60a}.

\subsection{Partially-extended (EP) sources} \label{sec:kr-ep}

The KR catalogue lists 41 of these sources. Table~\ref{tab:ep}
summarizes the CGPS view of this sample using the same notation as in
the previous tables.  One source (KR 145) appears to have been a
spurious object as there are no strong point sources or regions of
extended emission near the catalogued coordinates. 

Three of the sources have diameters greater than 11-arcmin.  KR 200 is
a large ($\sim 30$ arcmin) region of radio emission a portion of which is
directly associated with the optical \mbox{H\,{\sc ii}} region Sh
2-209.  KR 140 is a 12-arcmin scale \mbox{H\,{\sc ii}} region and KR
130 is the well-studied SNR 3C 58.

There are 13 sources which are not point sources but have diameters
$<9$ arcmin. Nine of these objects are radio sources associated with known
small-diameter optical \mbox{H\,{\sc ii}} regions and one is
associated with the nearby galaxy Maffei 2. 

%Table of Sources originally classified as EP in Kallas & Reich 1980

\begin{table*}
\centering
\begin{minipage}{140mm}
\caption{1420 MHz and 408 MHz data for EP sources}
\label{tab:ep}
\begin{tabular}{lccccccl}
\hline
KR & F$_\nu$ (1420) & $\sigma$ (1420) & F$_\nu$ (408)    & $\sigma$ (408) & $\alpha_{408}^{1420}$ & Diameter & Notes \\
   &   (mJy)        &       (mJy)     &    (mJy)         &     (mJy)      &                       &  \arcmin &       \\
\hline
13    & $1.13\times10^3$ & $2.6\times10^2$  &  $7.32\times10^2$  &  $9.4\times10^1$ &  $+0.3$  & 6  &  RRF 888; BFS 6      \\
15    & $3.10\times10^2$ & $9.3\times10^0$  &  $5.55\times10^2$  &  $1.7\times10^1$ &  $-0.5$  & 1  &  NVSS J212305+550027 \\
17    & $6.45\times10^2$ & $1.9\times10^1$  &  $5.46\times10^2$  &  $1.6\times10^1$ &  $+0.1$  & 2  &  RRF 899; Sh 2-187   \\
18    & $6.52\times10^2$ & $8.7\times10^0$  &  $4.30\times10^2$  &  $1.5\times10^1$ &  $+0.3$  & 6  &  RRF 929; BFS 8      \\
28A   & $2.56\times10^2$ & $7.9\times10^0$  &  $7.27\times10^2$  &  $2.3\times10^1$ &  $-0.8$  & 1  &  NVSS J213932+554030 \\
28B   & $1.64\times10^2$ & $5.3\times10^0$  &  $5.03\times10^2$  &  $1.7\times10^1$ &  $-0.9$  & 1  &  NVSS J213934+554445 \\
28C   & $5.32\times10^1$ & $2.5\times10^0$  &     $\cdots$       &
$\cdots$    & $\cdots$ & 1  &  NVSS J213943+554340 \\
\hline
\end{tabular}
\medskip
Table~\ref{tab:ep} is presented in its entirety in the electronic
edition of the journal.
\end{minipage}
\end{table*}

\begin{figure}
\includegraphics[width=84mm]{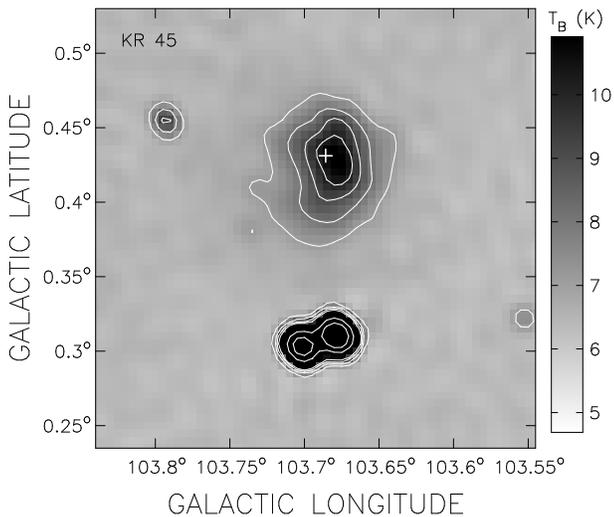}
%\vspace{3.5cm}
\caption{KR 45 at 1420 MHz. The original single source is actually a
  Galactic \mbox{H\,{\sc ii}} region and a pair of bright
  extragalactic sources. Contours are at 7, 8, 9, 10, 20, and
  30~K. The cross indicates the position of the infrared source IRAS
  22181+5716.}
\label{fig:kr45}
\end{figure}

\begin{figure*}
\centering
\begin{minipage}{140mm}
\includegraphics[width=140mm]{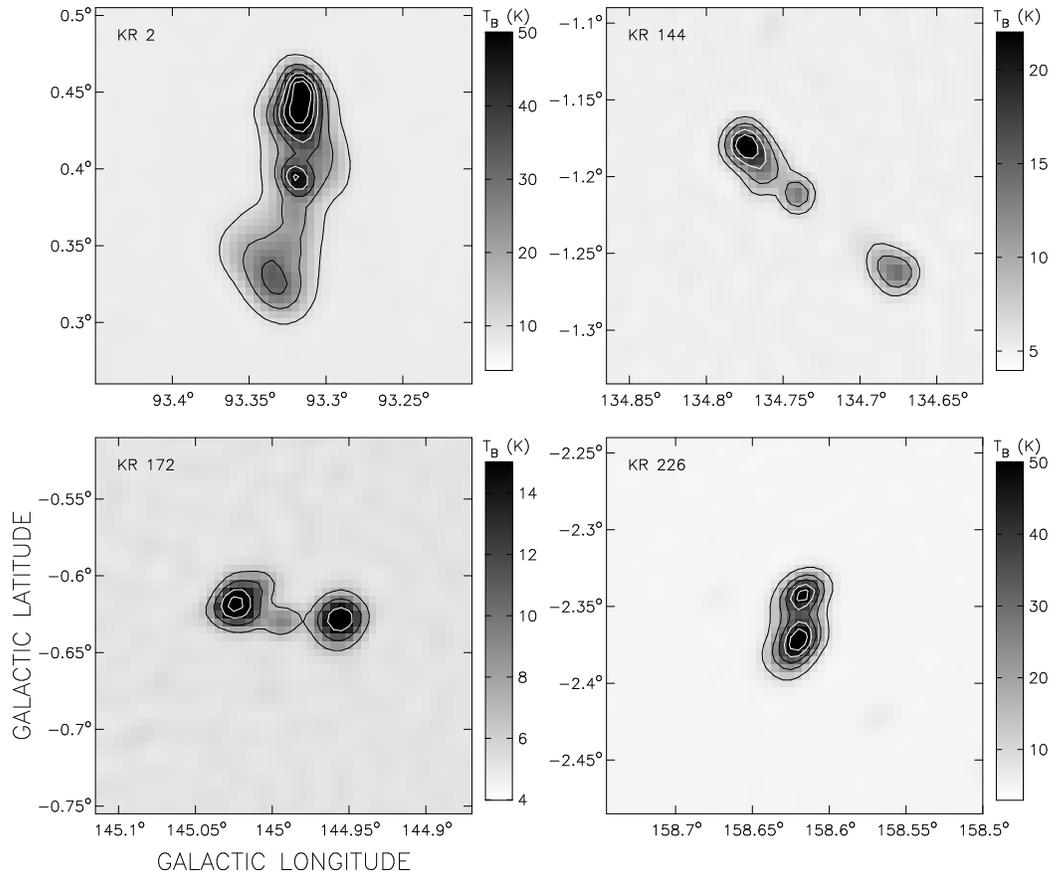}
%\vspace{3.5cm}
\caption{A sample of large radio galaxies at 1420 MHz. Each of these sources
  shows a distinct double radio lobe morphology with a compact or
  unresolved core. Contours for KR 2 and KR 226 are at 10 to 60~K at 10~K
  intervals; for KR 144 at 7, 10, 15 and 20~K; and for KR 172 at 7, 10, 13
  and 16~K.}
\label{fig:rgals}
\end{minipage}
\end{figure*}

KR 45 (RRF 981) is a combination of extended and point source emission
(see Figure~\ref{fig:kr45}). The extended radio emission is associated
with the distant \mbox{H\,{\sc ii}} region IRAS 22181+5716. Molecular line (CO)
observations towards this source detect emission at V$_\mathrm{LSR} = -63$
km~s$^{-1}$ placing the \mbox{H\,{\sc ii}} region at a heliocentric
distance of $\sim 7$~kpc. There is also a close double point source
(denoted 45A and 45B) which is unresolved in the lower resolution 408 MHz
images. These non-thermal point sources have no infrared counterparts
and are apparently just background extragalactic sources. The
remaining two extended objects (KR 144 and 172) both appear to be
radio galaxies with a distinct core/lobe morphology (see 
Figure~\ref{fig:rgals}). The objects shown in Figure~\ref{fig:rgals}
appear to be similar to the giant radio source WN 1626+5153 
discovered in the Westerbork Northern Sky Survey \citep{rot96}. 

Finally the remaining EP sources are all point sources at 1-arcmin
resolution. All but one (KR 58) are likely extragalactic sources
having a non-thermal spectral index and no detectable infrared emission.  KR 58
has an inverted spectrum and is the planetary nebula NGC 7354 
(IRAS 22384+6101).

\section{The nature of the point sources in the KR Catalogue} \label{sec:kr-p}

All of the KR point sources (135 in total) except one (KR 195) were
observed by the CGPS. Table~\ref{tab:p} summarizes the CGPS view of
this sample using the same notation as in the previous tables. 

The vast majority of these sources are point sources at 1-arcmin
resolution. As first demonstrated by \citet{fic86} most of these are
extragalactic sources as indicated in this study by their strongly negative
spectral index between 408 and 1420 MHz and lack of associated
infrared emission.

There are a few small extended sources in this subsample. KR 77, 212
and 228 are all regions of extended thermal emission associated with
the optical \mbox{H\,{\sc ii}} regions Sh 2-159, Sh 2-212 and Sh 2-217
respectively. Perhaps more interesting are the extended extragalactic
sources KR 2 and KR 226. Both of these objects are clearly radio
galaxies (see Figure~\ref{fig:rgals}) and were noted by
\citet{fic86} as being overresolved in his VLA images. KR 2 extends
for about 10-arcmin in its longest direction. Optical spectroscopy of
this source places it at a redshift of z=0.02 \citep{mas04}. KR 226
extends for about 5-arcmin and no studies of this object beyond
cataloging have been made. 

%Table of Sources originally classified as P in Kallas & Reich 1980 
 
\begin{table*} 
\centering 
\begin{minipage}{140mm} 
\caption{1420 MHz and 408 MHz data for P sources} 
\label{tab:p} 
\begin{tabular}{lccccccl} 
\hline
KR & F$_\nu$ (1420) & $\sigma$ (1420) & F$_\nu$ (408)    & $\sigma$ (408) & $\alpha_{408}^{1420}$ & Diameter & Notes \\
   &   (mJy)        &       (mJy)     &    (mJy)         &     (mJy)      &                       &  \arcmin &       \\
\hline 
2     & $2.87\times10^3$ & $8.6\times10^1$  &  $6.16\times10^3$  &  $1.8\times10^2$ &  $-0.6$  & 6  &   RRF 862   \\
5     & $4.37\times10^2$ & $1.3\times10^1$  &  $1.39\times10^3$  &  $4.2\times10^1$ &  $-0.9$  & 1  &   NVSS J213646+495318 \\
8     & $1.77\times10^3$ & $5.3\times10^1$  &  $1.07\times10^3$  &  $3.3\times10^1$ &  $+0.4$  & 1  &   NVSS J213701+510136 \\
9     & $3.22\times10^2$ & $9.9\times10^0$  &  $7.56\times10^2$  &  $2.4\times10^1$ &  $-0.7$  & 1  &   NVSS J213158+521415 \\
10    & $6.69\times10^2$ & $2.0\times10^1$  &  $1.24\times10^3$  &  $3.7\times10^1$ &  $-0.5$  & 1  &   NVSS J213340+521951 \\
11    & $7.72\times10^2$ & $2.3\times10^1$  &  $1.49\times10^3$  &  $4.5\times10^1$ &  $-0.5$  & 1  &   NVSS J213833+513550 \\
\hline
\end{tabular}
\medskip
Table~\ref{tab:p} is presented in its entirety in the electronic
edition of the journal.
\end{minipage}
\end{table*}

\begin{figure*}
\centering
\begin{minipage}{140mm}
\includegraphics[width=140mm]{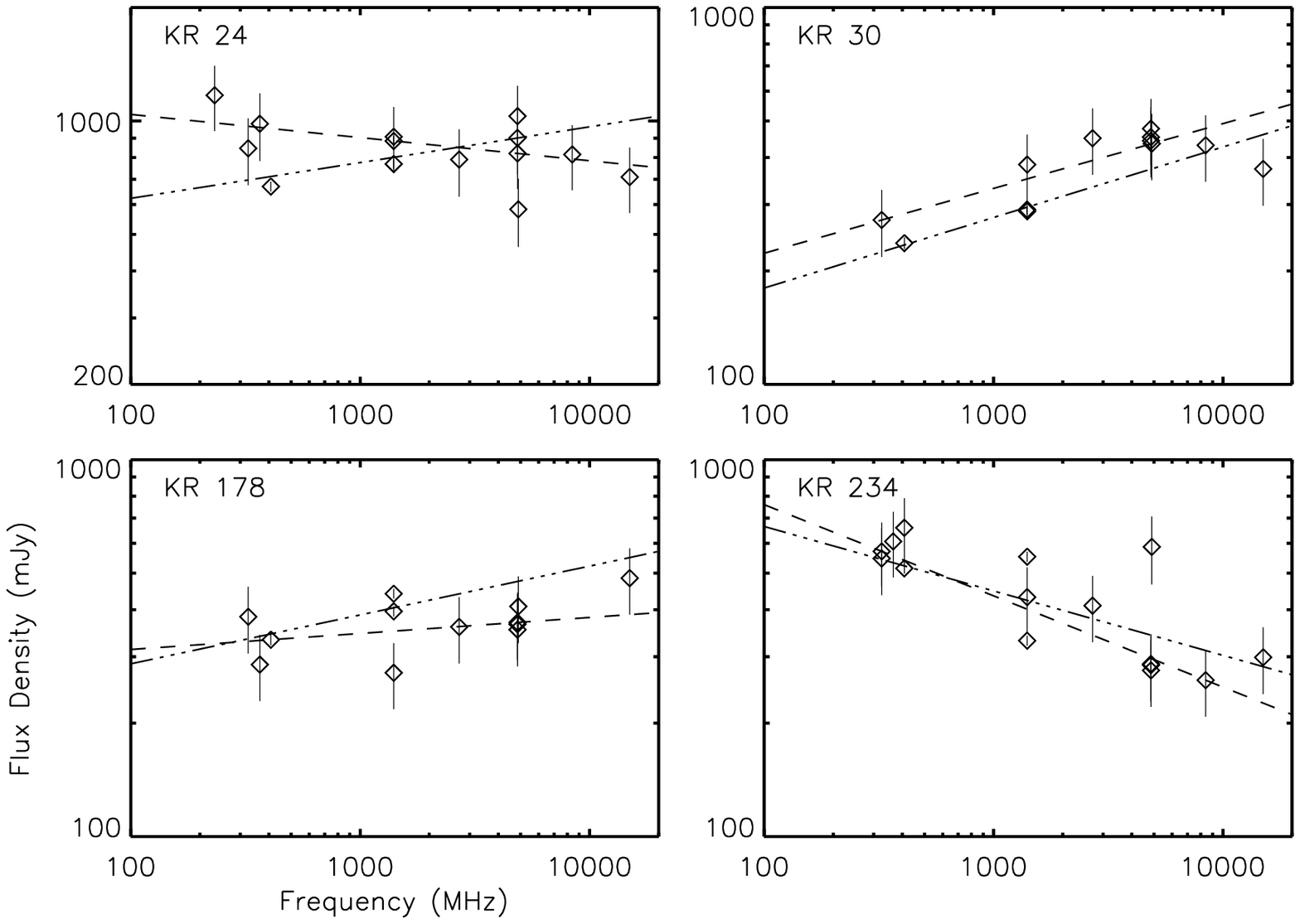}
%\vspace{3.5cm}
\caption{Radio spectra of flat spectrum KR sources with positive
  $\alpha_{408}^{1420}$ in the CGPS observations. Least absolute
  deviation fits to the data are shown with the dashed lines, and
  weighted least-squares fits are shown using the dot-dash lines.}
\label{fig:ps-flat-pos}
\end{minipage}
\end{figure*}

There are 14 flat spectrum sources of which three (KR 23, 208, and
212) are associated with optical \mbox{H\,{\sc ii}} regions (Sh 2-148,
Sh 2-211 and Sh 2-212 respectively). The remaining 11 sources have no
associated infrared emission and thus inferred to be extragalactic sources. We
examined the four flat spectrum sources with positive spectral indices in more
detail. CGPS data were combined with data from \citet{vol05} and
\citet{fic86} to obtain the spectra shown in Figure~\ref{fig:ps-flat-pos}. 

The radio spectrum of KR 24 is very flat over a wide frequency range,
and certainly flatter than expected just from 408 and 1420 MHz data. A
least absolute deviation fit to the data gives an overall spectral
index of $\alpha = -0.06$.  KR 178 is another very flat spectrum source with
least absolute deviation spectral index of $\alpha = +0.04$ over the
entire range of observations. KR 30 shows a slightly rising spectrum
with $\alpha = +0.2$. The highest frequency point suggests that the
spectrum may be flattening above 10 GHz. Finally the KR 234 radio
spectrum has a shallow negative slope spectrum of $\alpha = -0.2$. The
low frequency data points for KR 234 are in good agreement but there
is increased scatter at the higher frequencies. The large scatter observed
in the spectra of KR 24, 178 and 234 at particular wavelengths
suggests that these sources are variable. This is the likely reason that
the overall spectral index for these three sources is shallower
than the spectral index determined by the simultaneous CGPS
observations.

There are also eight inverted spectrum point sources.  Three of the
sources (KR 61, 67 and 72) are associated with optical \mbox{H\,{\sc
    ii}} regions (Sh 2-146, Sh 2-152 and Sh 2-156 respectively) and KR
138 is the compact \mbox{H\,{\sc ii}} region IRAS 02044+6031.
Molecular line emission at V$_\mathrm{LSR} \sim -55$ km~s$^{-1}$ has been
detected towards this \emph{IRAS} source placing it at a kinematic distance
of $\sim 5.5$~kpc. Unfortunately the velocity field model
of \citet{bra93} is quite uncertain around this longitude ($l\sim
130\degr$) for this velocity making corrections for streaming motions
problematic. Given its small angular size it it quite possible that KR
138 lies beyond the Perseus Arm. The remaining four sources have no
infrared counterpart and are most likely extragalactic. 

Such extragalactic radio sources with inverted spectra are interesting
because of the possibility that they are Gigahertz Peaked Spectrum
(GPS) sources. Astronomically these objects are of interest because
they may represent an early stage in the evolution of radio galaxies
\citep{ort06,ode98}. Observationally these objects are defined as
having a convex radio spectrum that peaks between 500 MHz and 10
GHz. The shape of the spectrum is most likely due to synchrotron
self-absorption \citep{ort06}. Below the peak frequency the average
spectral index is $0.51\pm0.03$ and above the peak it is $-0.73\pm0.06$
\citep{dev97}.  

\begin{figure*}
\centering
\begin{minipage}{140mm}
\includegraphics[width=140mm]{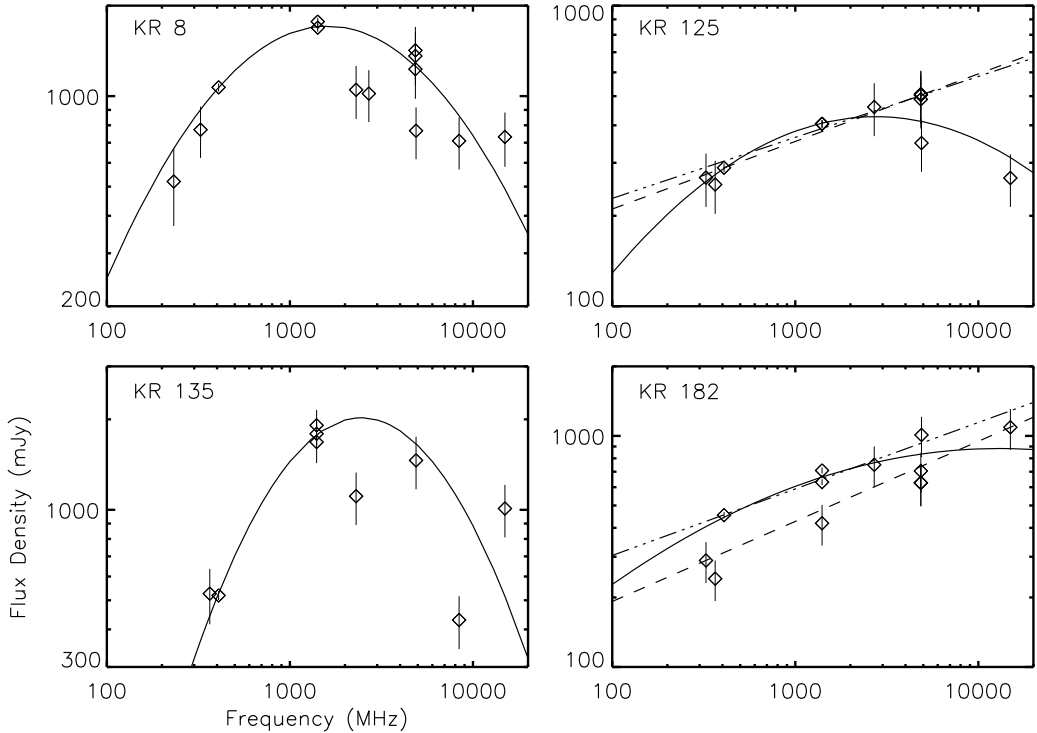}
%\vspace{3.5cm}
\caption{Radio spectra of KR sources with inverted spectra between 408
and 1420 MHz in the CGPS data. Second-order polynomial fits (see text
for details) are shown in each panel (solid line). For KR 125 and KR 182 linear
fits are also shown using the same style as in Figure~\ref{fig:ps-flat-pos}.}
\label{fig:kr-invert}
\end{minipage}
\end{figure*}

For each of the extragalactic inverted spectrum sources we combined
flux density measurements at other wavelengths from \citet{vol05} and
\citet{fic86} with the CGPS measurements. The spectra are shown in
Figure~\ref{fig:kr-invert}.  Following \citet{mar99} we fit a second
order polynomial of the form $ \log F_\nu = a + b \log \nu - c(\log \nu
)^2$. This curve is not physically motivated, rather it simply allows
us to easily identify sources with sufficiently high spectral
curvature. Sources with $c > 1.0$ have sufficient spectral curvature
to be considered GPS sources. 

\begin{figure*}
\centering
\begin{minipage}{140mm}
\includegraphics[width=140mm]{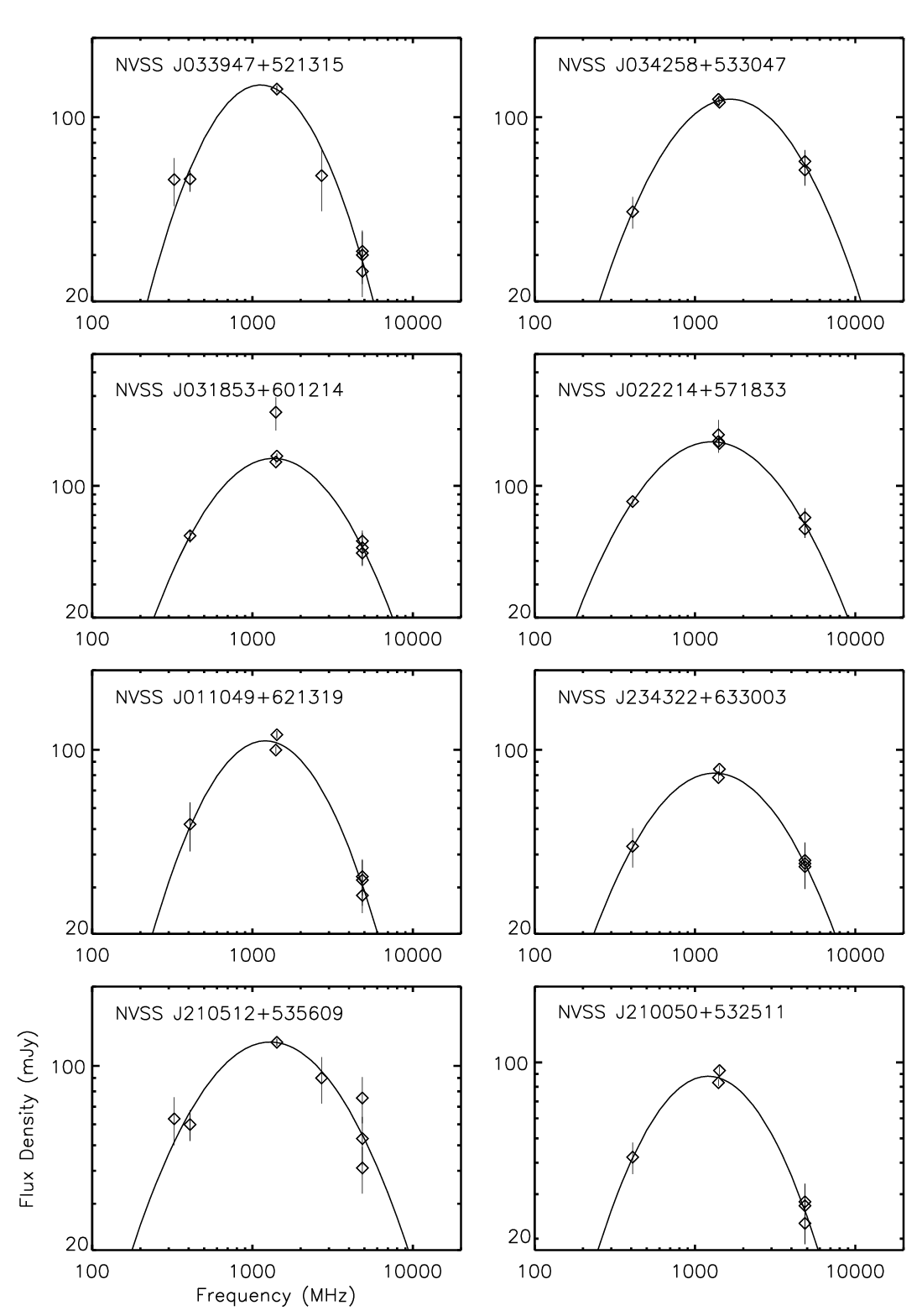}
%\vspace{3.5cm}
\caption{GPS Sources Candidates. All of these point sources have very
  high curvature radio spectra that peak in the GHz
  range. Second-order polynomial fits to the spectra are shown as
  solid lines and the NVSS designation is given in each panel.}
\label{fig:curve}
\end{minipage}
\end{figure*}

KR 8 does appear to have a convex spectra but the data above the peak
has a large amount of scatter and the curvature is not as high as one would
expect for a true GPS source ($c = 0.6$). 
KR 125 has a very low curvature spectrum ($c= 0.25$) with the
curvature arising almost entirely from the highest frequency data
point. Except for this point the spectrum is consistent with a rising spectrum
with $\alpha = +0.3$ from 300 to 4800 MHz.
KR 135 has a very steep low frequency spectral index and the cuvature
of the spectrum is quite high ($c=0.96$). Unfortunately the data
above the apparent peak in the spectrum are quite scattered and its
status as a GPS source is very uncertain.  
Finally, KR 182 shows a rising spectrum with $\alpha = +0.3$ with no
signs of any spectral curvature. There is a large amount of scatter in
the spectrum at both low and high frequency.

\begin{figure*}
\centering
\begin{minipage}{140mm}
\includegraphics[width=140mm]{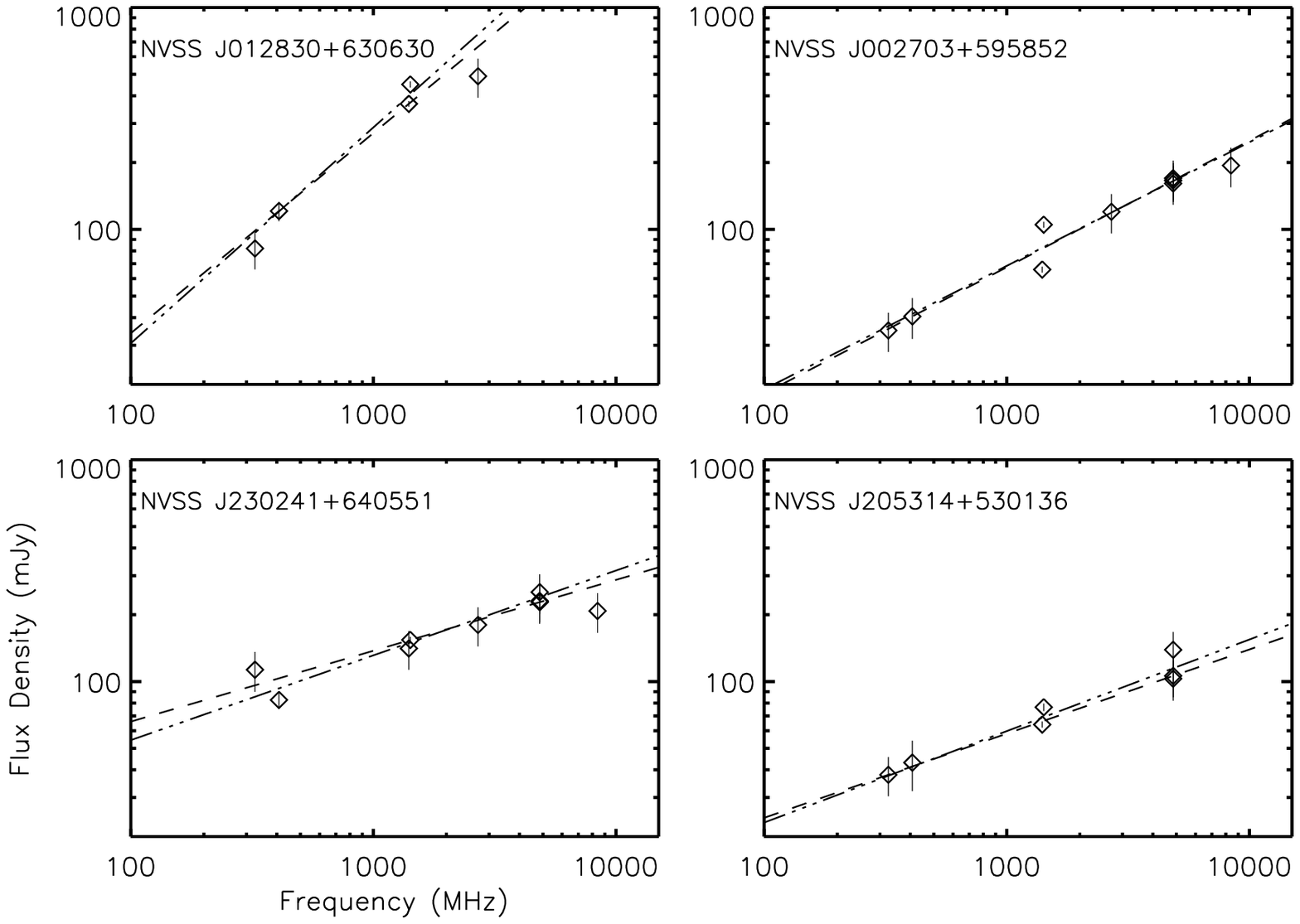}
%\vspace{3.5cm}
\caption{Rising spectrum sources. These objects are possible GPS
  sources with spectra peaking beyond 5 GHz. Linear fits to the data
  are shown in each case using the same style as in 
  Figure~\ref{fig:ps-flat-pos} and the NVSS designation is given in
  each panel.}
\label{fig:rise}
\end{minipage}
\end{figure*}
 
\section{Flat and Inverted-spectrum point sources} \label{sec:fiss}

The presence of extragalactic sources with both flat and inverted
spectra within the KR sample led us to examine all of the CGPS second
quadrant data for similar sources. To rapidly search for other point
sources with flat or inverted spectra the 1420 MHz images were first
convolved to the 408 MHz resolution. The brightness of the convolved
1420 MHz images were then scaled to the expected brightness at 408 MHz
assuming an optically thin thermal spectrum between 408 and 1420
MHz. The true 408 images were then subtracted from the scaled images
resulting in a series of difference images. Point sources with steep
negative spectral indices show up as distinct negative-valued sources
on the difference images thus allowing the rapid identification of
flat and inverted-spectrum sources. After candidate sources were
identified in this manner, flux densities were measured at 1420 and
408 MHz. Sources in the final sample had both measurable 408 flux
densities (complete to $\sim 50$ mJy at 408 MHz) and no visible
infrared emission in the ancillary CGPS infrared images.

Table~\ref{tab:if} shows the resulting sample of flat-spectrum and
inverted-spectrum sources. Column 1 gives the NVSS catalogue
designation, columns 2-5 give the flux density and error estimates at
1420 and 408 MHz, and column 6 gives the spectral index.

%Table of Inverted and Flat Spectrum Sources
%1-43  
\begin{table*} 
\centering 
\begin{minipage}{100mm} 
\caption{1420 MHz and 408 MHz data for Inverted and Flat Spectrum Sources} 
\label{tab:if} 
\begin{tabular}{lccccc} 
\hline
NVSS & F$_\nu$ (1420) & $\sigma$ (1420) & F$_\nu$ (408)    & $\sigma$ (408) & $\alpha_{408}^{1420}$ \\
     &   (mJy)        &       (mJy)     &    (mJy)         &     (mJy)      &                       \\
\hline 
J054044+391612 & $1.53\times10^2$ & $4.7\times10^0$ & $4.74\times10^1$ & $4.2\times10^0$ & $+0.9$ \\
J054052+372847 & $1.74\times10^2$ & $5.3\times10^0$ & $1.30\times10^2$ & $8.2\times10^0$ & $+0.2$ \\
J050905+352817 & $3.85\times10^2$ & $1.2\times10^1$ & $1.49\times10^2$ & $3.8\times10^1$ & $+0.8$ \\
J050920+385046 & $9.40\times10^1$ & $2.9\times10^0$ & $8.22\times10^1$ & $7.5\times10^0$ & $+0.1$ \\
J051346+400618 & $3.55\times10^2$ & $1.1\times10^1$ & $3.35\times10^2$ & $1.1\times10^1$ & $+0.0$ \\
J050948+395154 & $7.83\times10^1$ & $2.4\times10^0$ & $3.52\times10^1$ & $1.5\times10^0$ & $+0.6$ \\
\hline
\end{tabular}
\medskip
Table~\ref{tab:if} is presented in its entirety in the electronic edition of the journal.
\end{minipage}
\end{table*}

In order to identify potential GPS sources we examined in more detail
43 of the sources which had $\alpha_{408}^{1420} \geq +0.4$. As before,
radio data from the compilation of \citet{vol05} were used to
construct spectra over as wide a range of frequencies as
possible. Of these objects eight of them were found to have a curvature of $c >
+1$. The radio spectra of these objects are shown in Figure~\ref{fig:curve}.

We also found four other objects in the sample that had rising spectra
($\alpha \geq +0.3$ over the entire spectral range) combined with
little scatter (see Figure~\ref{fig:rise}). These sources may be
examples of, relatively rare, GPS sources with a peak above 5 GHz
similar to the point source 71P 52 (NVSS 213551+471022) examined by 
\citet{hig01}.

\section{Conclusions} \label{sec:conc}

The KR catalogue is very useful for Galactic studies as it contains
information on both compact and extended radio sources in the outer
Galaxy. Unfortunately the relatively low resolution of the survey
means that it overestimates the number of extended sources in the
outer Galaxy. This paper updates this catalogue based primarily on new higher
resolution images of the outer Galaxy at 1420 MHz obtained as part of
the CGPS. We have clearly identified sources that were misclassified
as extended objects and have determined which sources remain
unresolved at 1-arcmin scale resolution. The simultaneous 408 MHz CGPS 
observations, combined with ancillary infrared data, also have
allowed the nature of all of the observed KR sources to
be determined with some confidence. Attention has been drawn
particularly to a large number of unstudied Perseus Arm \mbox{H\,{\sc
    ii}} regions
(including the extremely large KR 1 complex), objects previously
considered to be SNR candidates (e.g., KR 171), and a sample of large
radio galaxies (e.g., KR 144).

In addition, through the examination of the 408 and 1420 MHz
CGPS images, this study has identified a sample of flat-spectrum and
inverted-spectrum extragalactic radio sources based upon their 408 and
1420 MHz flux densities. A subset of these objects was examined in
more detail and a new sample of GPS sources has been compiled.

\section*{Acknowledgments}
I would like to thank ISU undergraduate students Jason Murphy and Jon Patterson
for their assistance on this project. The Dominion Radio Astrophysical
Observatory is operated by the National Research Council of
Canada. The Canadian Galactic Plane Survey is supported by a grant
from Natural Science and Engineering Research Council of Canada.

\end{document}